\documentclass[useAMS,usenatbib,usegraphicx]{mn2e}
\usepackage{times}
\usepackage{txfonts}
\usepackage{ulem}
\bibpunct{(}{)}{;}{a}{}{,}
\usepackage{xspace} 
\usepackage{setspace}

\newcommand {\x} {\textit{XMM-Newton}\xspace}
\newcommand {\h} {H.E.S.S.\xspace}
\newcommand {\snr} {G22.7$-$0.2\xspace}
\newcommand {\src} {HESS J1832$-$093\xspace}
\newcommand {\radecpos} {$\rm RA=18^h 32^m 50^s \pm 3^s_{stat} \pm 2^s_{syst}, \rm  Dec=-9^\circ 22' 36'' \pm 32''_{stat} ~\pm 20''_{syst}~(J2000)$\xspace}
\newcommand {\vheindex} {$\Gamma = 2.6 \pm 0.3_{\rm stat} \pm 0.1_{\rm syst}$\xspace}
\newcommand {\brem}{bremsstrahlung\xspace}
\newcommand {\flux}{$\Phi_0=(4.8 \pm 0.8_{\rm stat}\pm 1.0_{\rm syst})\,\times\,10^{-13}\,\rm{cm} ^{-2}\,s^{-1}\,TeV^{-1}$}
\newcommand {\inteflux}{$\rm I(\rm E> \rm 1\,TeV)=(3.0\pm0.8_{\rm stat}\pm0.6_{\rm syst})\,\times 10^{-13}\,\rm cm ^{-2}\,s^{-1}$} 
\newcommand {\xmmu}{XMMU J183245$-$0921539\xspace}
\newcommand {\irmass}{2MASS J18324516$-$0921545\xspace}
\newcommand {\mj}{$m_J=15.52 \pm 0.06~\rm mag$\xspace}
\newcommand {\mh}{$m_H=13.26 \pm 0.04~\rm mag$\xspace}
\newcommand {\mk}{$m_K=12.17 \pm 0.02~\rm mag$\xspace}

\title[Discovery of the VHE gamma-ray source \src]{Discovery of the VHE gamma-ray source \src in the vicinity of SNR \snr}

\author[H.E.S.S. Collaboration]{\normalsize H.E.S.S. Collaboration,
 A.~Abramowski,$^{1}$
 F.~Acero,$^{41,*}$
 F.~Aharonian,$^{2,3,4}$
 F.~Ait Benkhali,$^{2}$
 A.G.~Akhperjanian,$^{5,4}$
 E.~Ang\"uner,$^{6}$
 G.~Anton,$^{7}$
 \newauthor \normalsize 
 S.~Balenderan,$^{8}$
% ...
\\
\\
$^1$
Universit\"at Hamburg, Institut f\"ur Experimentalphysik, Luruper Chaussee 149, D 22761 Hamburg, Germany\\
$^2$
Max-Planck-Institut f\"ur Kernphysik, P.O. Box 103980, D 69029 Heidelberg, Germany\\
$^3$
Dublin Institute for Advanced Studies, 31 Fitzwilliam Place, Dublin 2, Ireland\\
$^4$
National Academy of Sciences of the Republic of Armenia, Yerevan\\
$^5$
Yerevan Physics Institute, 2 Alikhanian Brothers St., 375036 Yerevan, Armenia\\
$^6$
Institut f\"ur Physik, Humboldt-Universit\"at zu Berlin, Newtonstr. 15, D 12489 Berlin, Germany\\
$^7$
Universit\"at Erlangen-N\"urnberg, Physikalisches Institut, Erwin-Rommel-Str. 1, D 91058 Erlangen, Germany\\
$^8$
University of Durham, Department of Physics, South Road, Durham DH1 3LE, U.K.\\
% ...
$^{41}$
Laboratoire AIM, CEA-IRFU/CNRS/Universit\'e Paris Diderot, Service d'Astrophysique, CEA Saclay, 91191 Gif sur Yvette, France\\
% ...
$^*$
E-mail: laffon@cenbg.in2p3.fr (H\'el\`ene Laffon); fabio.acero@cea.fr (Fabio Acero)\\
}

\begin{document}

\date{Accepted 2014 XX XX. Received 2014 May XX XX; in original form 2013 November 25
  } \pagerange{\pageref{firstpage}--\pageref{lastpage}}
\pubyear{2014}

\maketitle

\label{firstpage}

\setstretch{1}

%
%
\begin{abstract}
\small The region around the supernova remnant (SNR) W41 contains several TeV sources and has prompted the \h Collaboration to perform deep observations of 
this field of view. This resulted in the discovery of the new very high energy (VHE) source \src, at the position \radecpos, spatially coincident with a part of the radio shell of the neighboring remnant \snr.  
The photon spectrum is well described by a power-law of index \vheindex and a normalization at 1 TeV of \flux. 
The location of the gamma-ray emission on the edge of the SNR rim first suggested a signature of escaping cosmic-rays illuminating a nearby molecular cloud.
Then a dedicated \x observation led to the discovery of a new X-ray point source spatially coincident with the TeV excess. Two other scenarios were hence proposed to
identify the nature of \src. Gamma-rays from inverse Compton radiation in the framework of a pulsar wind nebula scenario or the possibility of gamma-ray production within a binary system are therefore also considered.
Deeper multi-wavelength observations will help to shed new light on this intriguing VHE source. 
\end{abstract}

\begin{keywords}
astroparticle physics - gamma rays: general -  ISM: individual objects: \src -    ISM: individual objects: SNR \snr
\end{keywords}

\maketitle

\section{Introduction}

\h (High Energy Stereoscopic System) is an array of five imaging atmospheric Cherenkov telescopes located 
$1800~ \rm m$ above sea level in the Khomas Highland of Namibia. The first four telescopes have been fully operational since 2004 \citep{crab}, 
while the fifth telescope started operation in September 2012. 
The \h Collaboration has been conducting a systematic scan of the Galactic plane, which led to the discovery of a rich population 
of very high energy (VHE, E$\ge$100 GeV) gamma-ray sources.
The majority of these galactic sources are extended beyond the \h point spread function (PSF), which is of the order of 6$'$,
and mostly comprise supernova remnants (SNRs) and evolved pulsar wind nebulae (PWNe). 
Point-like sources are also observed in the Galactic plane and are generally associated with gamma-ray binaries (e.g. LS 5039, \citet{Aharonian_LS5039})
and with young PWNe such as G0.9$+$0.1 \citep{Aharonian_G0.9}. 
Furthermore, in the particular case of HESS J1943$+$213, an identification of the VHE point-like source in the Galactic 
plane with a background active galactic nucleus (AGN) is currently the most likely hypothesis \citep{hessj1943}.

The paper at hand deals with the field of view around SNR \snr, which is close to SNR W41 in sky projection. The discovery of a new point-like TeV source, \src, 
is reported in Section \ref{gamma} as well as the search for a GeV counterpart with the \textit{Fermi}-LAT. 
The TeV emission lies close (about 1$'$ away) to the radio rim of the supernova remnant \snr.
This SNR shows a non-thermal ring of 26$'$ diameter in radio \citep{Shaver}
and partially overlaps the neighboring remnant W41.
However, there is no obvious flux enhancement in the radio data around the position of \src.
Using the $\Sigma$-D relation given by \citet{guseinova} which connects the surface brightness $\Sigma$ of a supernova remnant 
with its diameter D, the estimated distance to \snr is approximately $(3.7\pm1.1)\rm~kpc$ \citep{guseinovb}.
The source location at the edge of the SNR shell could suggest a signature of escaping hadronic cosmic-rays which would illuminate 
dense material such as molecular clouds. Such scenario is considered a prime opportunity to unambiguously study hadronic cosmic-rays accelerated in supernova remnants. 
This possibility is examined in Section \ref{hadro}.
However, the compact nature of the TeV emission together with the detection of a new X-ray counterpart is at odds with this scenario. Dedicated \x data at the position of \src have led to the discovery 
of this new potential counterpart, the X-ray point source \xmmu, as detailed in Section \ref{compact}. 
As a consequence, two scenarios of compact objects, a young PWN or a binary system origin, are considered in Section \ref{compact} in order to explain the VHE emission.

\section{Multi-Wavelength data analysis}\label{gamma}

\subsection{\h}\label{hess}

A standard analysis method with Hillas event reconstruction \citep{crab} is adopted to study the field of view of interest.
A multi-variate analysis is used \citep{parismva} to provide improved discrimination between 
hadrons and gamma-rays. A minimum charge cut of 110 photo-electrons in the shower images is applied to the data, resulting in an energy threshold of about 450 GeV.

\begin{figure}
\includegraphics[width=92mm,bb=25 30 675 624,clip]{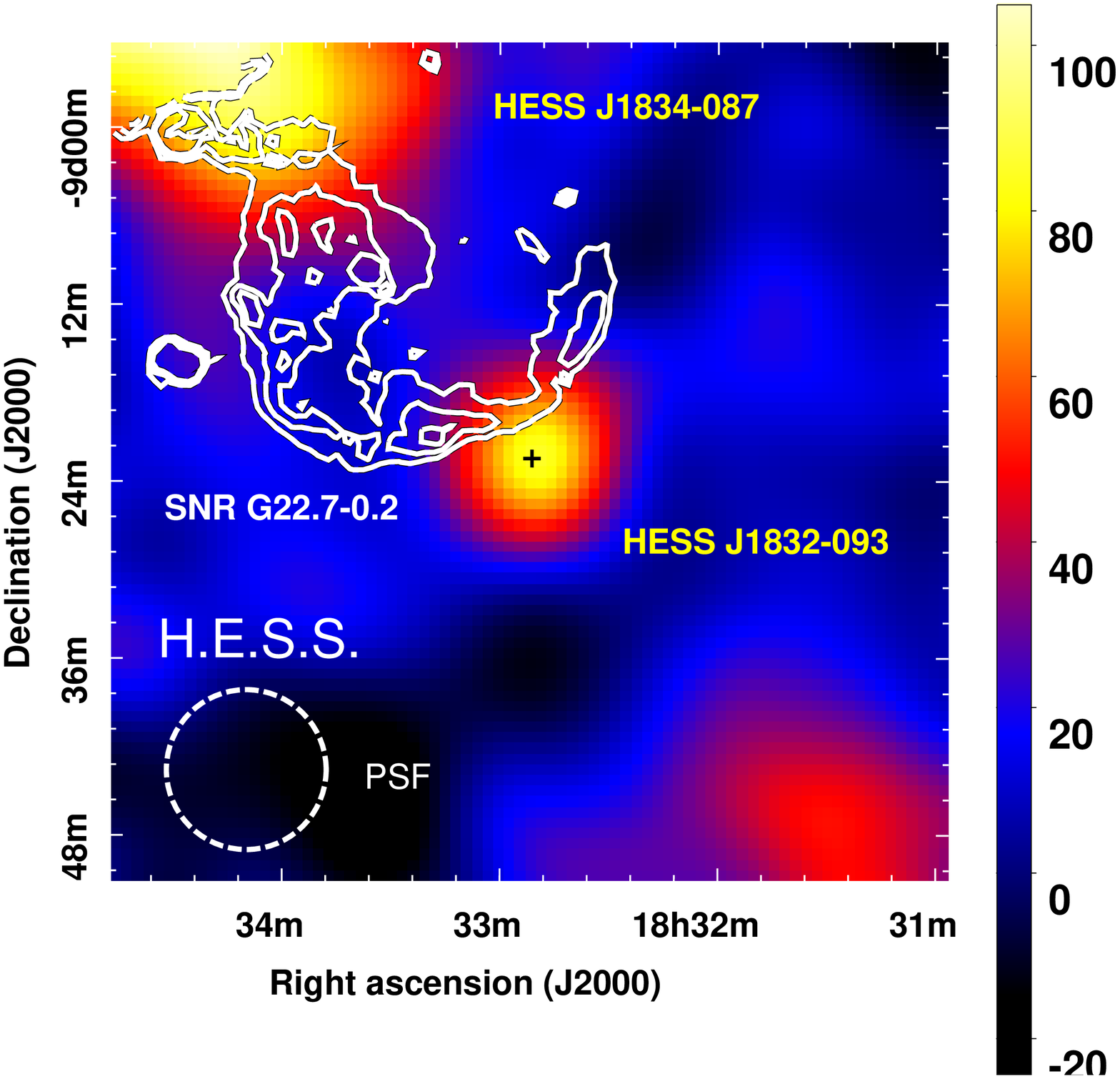}
\centering
\caption{\h excess map smoothed with a 2D-gaussian of width corresponding to the r$_{68}$ value of $0.081^\circ$ (represented by the dashed circle). Units are counts per integration area. The best-fit position of \src with statistical errors is represented by the black cross. The SNR  \snr observed at 1.4 GHz  \citep{magpis} is represented by the white contours. The emission seen on the upper left is a small part of HESS J1834$-$087 \citep{hess2006survey}, the TeV source in spatial coincidence with SNR W41. 
}
\label{HESSexcess}
\end{figure}

A standard run selection procedure is used to remove bad quality observations in order to study the newly discovered source. 
The available data set in this region covers a zenith angle ranging between 13$^\circ$ and 50$^\circ$ (mean value of 25$^\circ$) and comprises 67 hours live time of observations, taken from 2004 to 2011.
Using this data set, the new source, named \src, is detected with a peak significance of $7.9~\sigma$ pre-trials, corresponding to a post-trial detection significance of $5.6~\sigma$.
The average angular resolution (r$_{68}$) obtained for the selected data set is $0.081^\circ$ 
at the source position.
The excess map of the field of view centered on the new detected source and smoothed with the r$_{68}$ value is presented in Fig. \ref{HESSexcess}.

A two-dimensional symmetrical Gaussian function is used to determine the position and size of the TeV emission with 
a $\chi^2$ minimization. 
The best-fit position is \radecpos ($\chi^2$/ndf=0.89).
No significant extension was found for the source and an upper limit of 0.074$^\circ$ at a $99\%$ confidence level (C.L.) is derived.

In order to broaden the accessible energy range
the charge cut of the shower images is lowered to a minimum of 80 photo-electrons, resulting in an energy threshold of $\sim$400 GeV.
The forward-folding method described in \citet{crab} is applied to the data to derive the spectrum.
Source counts are extracted from a circular region of 0.1$^\circ$ radius around the best fit position of \src, a size optimized for point source studies with the applied cuts \citep{parismva}.

\begin{figure}
\centering
\includegraphics[width=95mm]{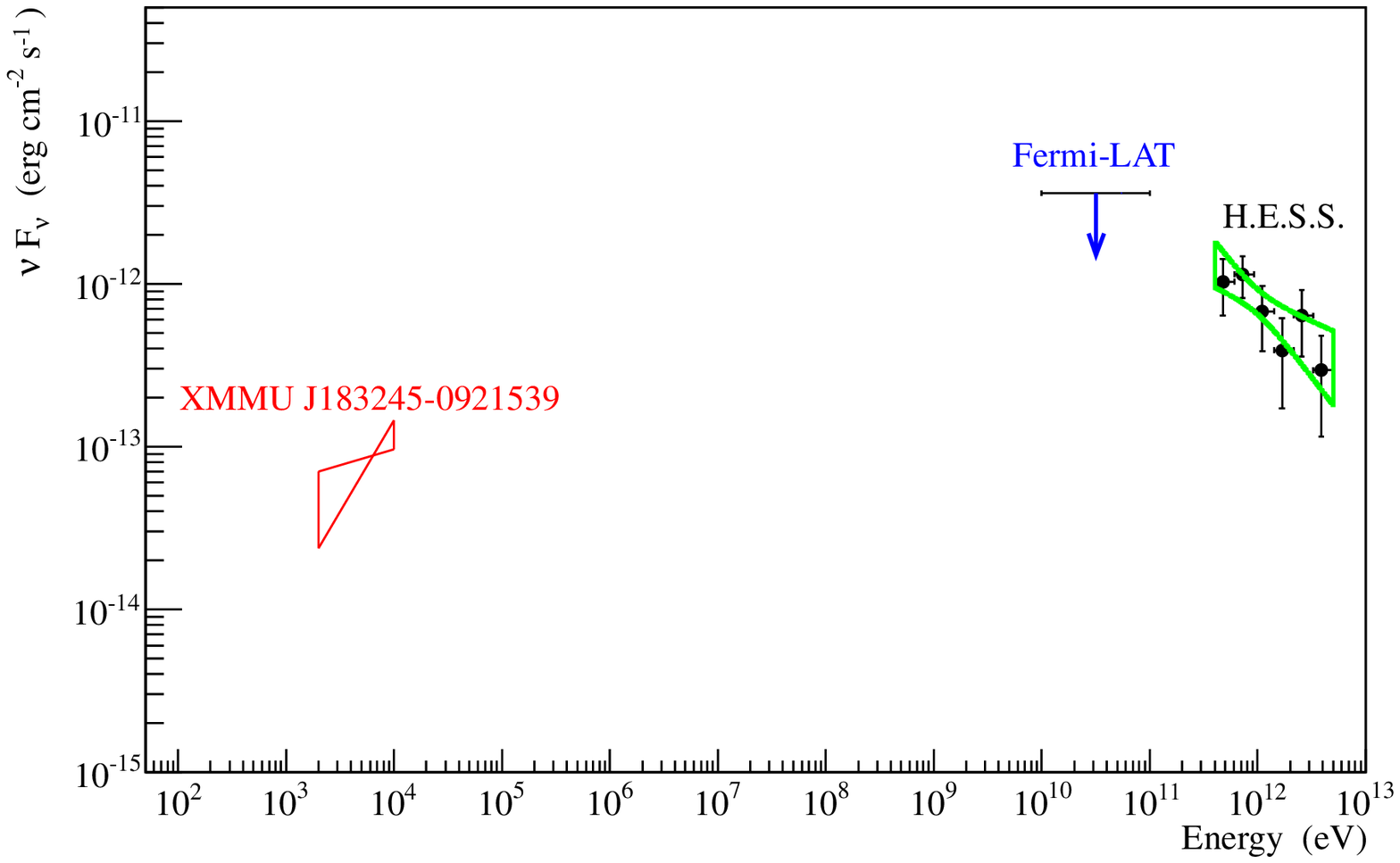}
\caption{
Spectral energy distribution from the region of VHE emission from \src.
The VHE gamma-ray spectrum observed with \h is displayed together with the upper limit obtained between 10 and 100 GeV with the Fermi-LAT. The green contour represents the 1$~\sigma$ confidence level of the fitted spectrum using a power-law hypothesis. Only statistical errors (68$\%$ confidence level) are shown for the spectral points. The X-ray power-law model of \xmmu is overlaid in red, taking into account the statistical uncertainties only.
}
\label{spectrum}
\end{figure}

The spectrum obtained between 400 GeV and 5 TeV (displayed in Fig. \ref{spectrum}) is well described by a power-law (PL)
${{d\Phi} \over {dE}}\,=\,\Phi_0 \left( {{E} \over {1\,\rm TeV}} \right) ^{-\Gamma}$,
with an index \vheindex and a differential flux normalization at 1 TeV of \flux.
The integrated flux above $1\rm~TeV$ is \inteflux, corresponding roughly to $1\%$ of the Crab Nebula flux above the same energy \citep{crab}.

A search for curvature in the gamma-ray spectrum was performed by fitting log-parabola and exponential cutoff power-law models
to the data.
While not ruled out, these models are not favored since the improvement in fit quality compared to the simple power-law model is not statistically significant.

Long-term light curves were produced with different integrated times (run, night and month) and Z-transformed discrete correlation functions \citep{Alexander1997} 
were applied to the data to look for periodicity. However no significant temporal variability was detected in the \h data set.

\subsection{\textit{Fermi}-LAT}

The Fermi Large Area Telescope (LAT) is a gamma-ray telescope operating in the 20 MeV to 300 GeV energy range \citep{FermiLAT}.
No \textit{Fermi}-LAT source is listed at the position of \src in the Fermi 2-year catalog \citep[2FGL, ][]{Fermi2FGL}.
However, this field of view located close to the Galactic plane is very rich in gamma-ray sources and diffuse emission, 
making the analysis challenging.
Furthermore, the \textit{Fermi} source 2FGL J1834.3-0848, in spatial coincidence with SNR W41, 
lies very close to \src and the \textit{Fermi} angular resolution at low energy does not allow
different potential contributions to this source to be distinguished.
Therefore a dedicated analysis was performed in the field of view to look for a potential counterpart in the GeV range.
The analysis was carried out with 4 years of data and above 10 GeV as a compromise between statistics and background from the diffuse Galactic emission (dominating for $E < 10$ GeV).
The Instrument Response Functions (IRFs) $\rm P7SOURCE\_V6$  and the \textit{Source} class events were used \citep[see][ for details about the event classification and IRFs]{FermiIRF}. 
The corresponding Galactic diffuse background ($\rm gal\_2yearp7v6\_v0.fits$) and the extragalactic isotropic background ($\rm iso\_p7v6source.txt$),
distributed with the  Fermi Science Tools\footnote{http://fermi.gsfc.nasa.gov/ssc/data/analysis/}, were used. 
In addition to the diffuse backgrounds, a model of the nearby sources within a 5$^\circ$ radius was built based on the 2FGL catalog \citep{Fermi2FGL}. 
No significant gamma-ray excess is found on top of the model previously built. 
An energy flux upper limit of $3.6\times10^{-12} \rm \, erg \,cm^{-2} \,s^{-1}$ in the 10-100 GeV band is then obtained at 99$\%$ C.L., assuming a point source at the position of \src. This upper limit is shown on the spectral energy distribution (SED) displayed on Fig. \ref{spectrum}.

\subsection{\textit{XMM-Newton}}\label{xmm}

In order to constrain the nature of the source \src, a dedicated observation with the X-ray \textit{XMM-Newton} satellite was performed 
in March 2011 for 17$\rm~ks$.
After filtering out proton flare contamination, 13$\rm~ks$  and 7$\rm~ks$ of exposure time remained for the two EPIC-MOS cameras and for the EPIC-pn camera, respectively. 
The data were processed using the \textit{XMM-Newton} Science Analysis System (v10.0). The instrumental background
 was derived from a compilation of blank sky observations \citep{carter07}, renormalized to the actual exposure using 
 the count rate in the 10-12 keV energy band.

\begin{figure}
\centering
\includegraphics[width=85mm,bb=52 65 508 530,clip]{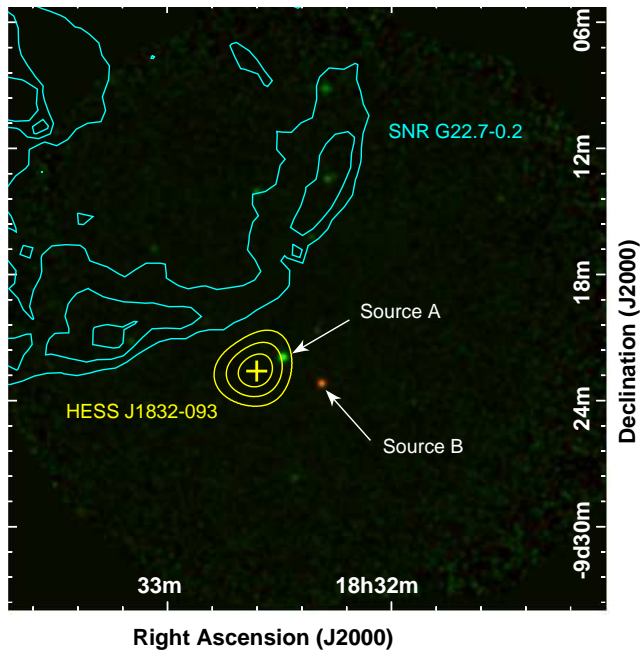}
\caption{
XMM-Newton composite flux map of the field of view around \src in the 0.5-2$\rm~keV$ (red) and 2-6$\rm~keV$ (green) energy ranges.
The SNR \snr observed at 1.4 GHz is overlaid in cyan contours.
The yellow cross symbolizes the best-fit position of \src with corresponding errors. The confidence contour levels (68$\%$, 95$\%$ and 99$\%$) of the source position fit are also shown in yellow. Two point-like sources are detected near the position of \src: Sources A and B discussed in Section \ref{xmm}. 
No diffuse emission from the SNR shell segment is seen in the X-ray data.
The maximal flux values are 3.5 $\times 10^{-4}\,$cm$^{-2}\,$s$^{-1}$arcmin$^{-2}$ and 3.6 $\times 10^{-4}\,$cm$^{-2}\,$s$^{-1}$arcmin$^{-2}$ for the red and green maps, respectively.
}
\label{xraymap}
\end{figure}

The brightest object in the \textit{XMM-Newton} field of view is a point-like source (Source A in Fig. \ref{xraymap}) located at 
$\rm RA=18^h32^m45^s.04, \rm Dec=-9^\circ~21'~53''.9 $ with a 90$\%$ C.L. error radius of 2.3$''$ which is  
1.5$'$ away from the best-fit position of the \h excess. This new source, named \xmmu, is located
within the 99\% C.L. contours of the \h best fit position and is a potential counterpart to the VHE source.
Other point sources detected in the field of view are either too soft (such as Source B, Fig. \ref{xraymap}) or too far away from \src to be considered as potential counterparts.

Spectra from the three instruments were extracted from a 15$''$ radius circular 
region centered on \xmmu. Both an absorbed power-law model and an absorbed black body model were tested.
The best fit parameters for the power-law model are a column density
N${_{\rm H}}= 10.5^{+3.1}_{-2.7} \times 10^{22}\,$cm$^{-2}$, a photon index $\Gamma=1.3^{+0.5}_{-0.4}$ 
and an unabsorbed energy flux $\Phi$(2-10 keV)$= 6.9^{+1.7}_{-2.8} \times 10^{-13}\,$erg$\,$cm$^{-2}\,$s$^{-1}$,
with a p-value\footnote{The p-value corresponds to the null-hypothesis probability} of 0.75. 
The absorbed black-body fit yields N${_{\rm H}}= 5.5^{+1.3}_{-1.8} \times 10^{22}\,$cm$^{-2}$, a temperature $\rm kT=1.9^{+0.3}_{-0.2}\,$keV,  
an unabsorbed energy flux $\Phi$(2-10 keV)$= 5.7 ^{+1.3}_{-2.2} \times 10^{-13}\,$erg$\,$cm$^{-2}\,$s$^{-1}$, and a p-value of 0.73.
Given the low level of statistics, no spectral model can be rejected, as shown by the p-values.
However, the fitted temperature of the black-body model is much higher than usually observed 
for cooling neutron stars ($\sim \,$0.2$\,$keV) or central compact objects (CCOs; $\sim \,$0.5$\,$keV).
Such a high temperature can be observed in bursting binary systems, but due to the lack of evidence of bursting behavior 
in X-rays, this scenario is not considered in the following discussion.
Hence, the power-law model is adopted to characterise the X-ray emission of \xmmu and it is displayed on Fig. \ref{spectrum}.

Because of the low statistics and the fact that the \textit{XMM-Newton}  observation was performed in imaging mode
(timing resolutions of 2.6 s and 73 ms for MOS and pn instruments), no detailed pulsation search could be carried out.
Future deeper observations in timing mode could provide better constraints on the nature of this X-ray source.

A comparison of the absorption along the line of sight obtained from the X-ray spectral model with the 
column depth derived from the  atomic (HI) and molecular ($^{12}$CO, J=1$\rightarrow$0 transition line) gas 
can be used to provide a lower limit on the distance to \xmmu, as described in \citet{HESSJ1731}.
A minimal distance of  $\sim$5 kpc is thus derived using the lower bound of the fitted N$_{\rm H}$ obtained with the power-law model.

\section{A hadronic origin?}\label{hadro}

\begin{figure}

\centering
\includegraphics[width=85mm]{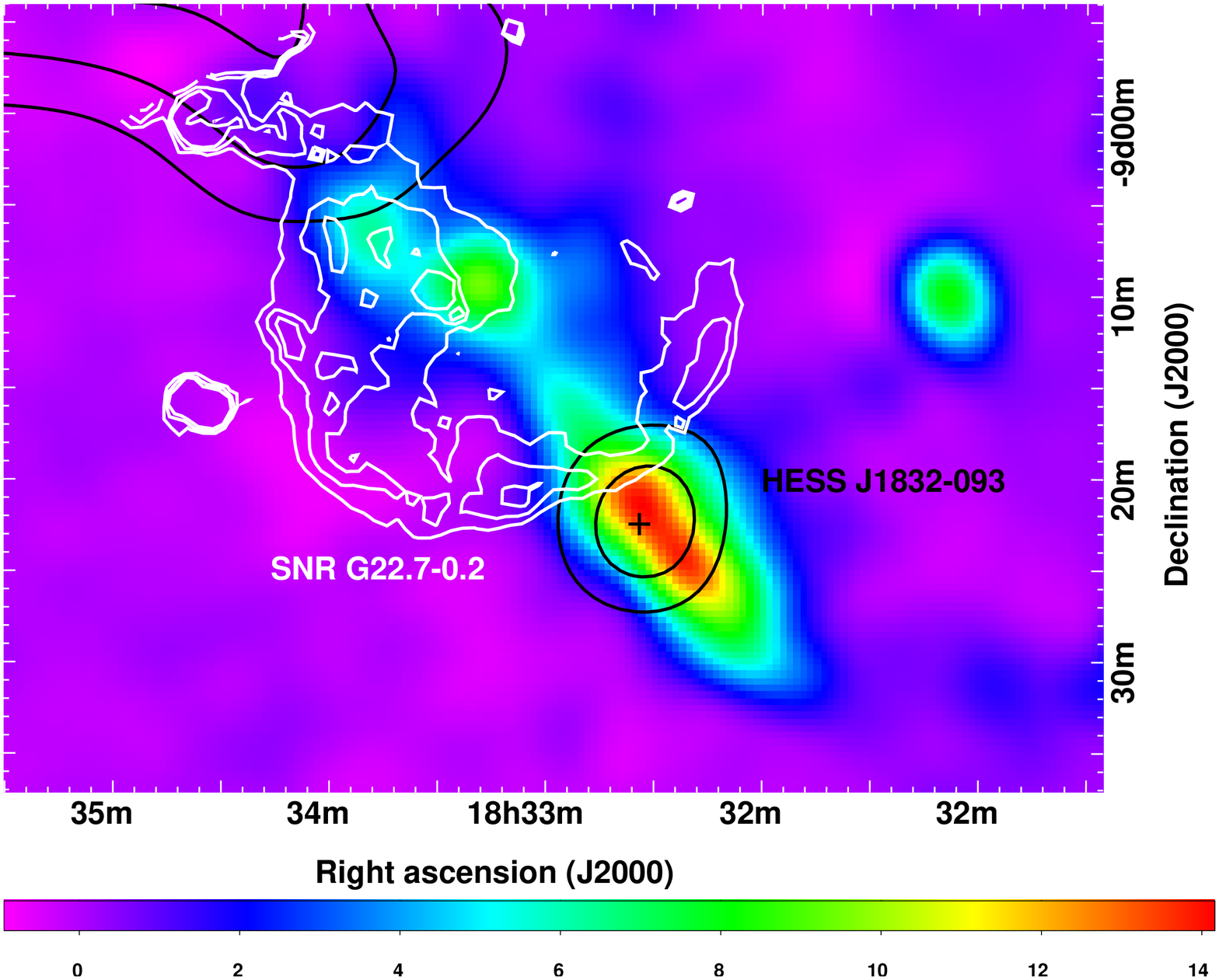}
\caption{Integrated $^{13}$CO antenna temperature in arbitrary units \citep{grs} in a velocity range of 26 to 31 km s$^{-1}$ smoothed with the average \h PSF for this data set. The gamma-ray excess of  Fig. \ref{HESSexcess} is shown in black contours (50, 75 and 100 gamma levels) while the radio observation of SNR \snr \citep{magpis} at 1.4 GHz is overlaid in white contours.
(0.002 and 0.005 mJy/beam). 
The black cross represents the best-fit position of the \h excess with corresponding errors.}
\label{clouds}
\end{figure}

The Galactic Ring Survey (GRS) performed with the Boston University FCRAO telescopes \citep{grs} provides measurements 
of the $^{13}$CO (J=1$\rightarrow$0) transition line covering the velocity range from -5 to 135 km s$^{-1}$ in this 
region. The detection of this line is evidence for the presence of dense molecular clouds (MCs) that are known to be targets for 
cosmic-rays and hence gamma-ray emitters via neutral pion production and decay.
Several MCs measured at different radial velocities are found around the source \src.
The molecular structure showing the best spatial coincidence with the TeV emission is selected and shown in Fig. \ref{clouds}. 
The MC near distance of about $2.3~\rm kpc$ given by the Galactic rotation curve model provided by \citet{hou09} is compatible with the distance estimate to the remnant.
Following the approach described in \citet{simon}, the integrated antenna temperature on the MC velocity range is used to derive the gas mass of the structure which is $\sim 700~\rm M_{\odot}$, corresponding to a gas density of  $\sim 20~\rm cm^{-3}$. 

TeV emission from the direction of \snr might be related to protons either coming from the cosmic-ray (CR) sea or accelerated in early phases of a nearby SNR and interacting in dense molecular structures, producing neutral pions that decay into gamma rays.
This scenario has already been invoked e.g. to explain ``dark'' TeV sources (e.g. \citet{Gabici2009}).
For this hypothesis to work, localised high density target material is needed in order to explain that only a very small fraction of \snr emits gamma-rays.
$^{13}$CO measurements show the presence of such structures near \src, as seen on Fig. \ref{clouds}.
Although no significant extension of the TeV source is detected, the upper limit of 0.074$^\circ$ is consistent with a slightly extended emission region as may be expected from the MC spatially coincident with \src.
The expected \brem emission from accelerated electrons can be neglected compared to the hadronic contribution since the proton to electron ratio p/e should be $\gg100$ for multi-TeV energies 
(e.g. \citet{Yuan2012} and references therein).

Using the mass and distance of the selected MC and following Eq. 10 of \citet{Aharonian1991}, the CR density enhancement factor $k_{CR}$ can be estimated in units of the local CR density, corresponding to a value of 780.
 Such a high enhancement factor require the presence of a nearby CR source
 such as SNR \snr in order to explain the observed TeV emission.
Moreover,  \citet{Aharonian1996} show that, given the SNR radius of about 10 pc and an assumed age around 10$^{3}$ yr, such a high $k_{CR}$ value is expected for a slow diffusion coefficient of $D\sim10^{27}\rm~cm^2~s^{-1}$ for 10 TeV hadrons, but excluded for a diffusion coefficient of $D \sim10^{29}\rm~cm^2~s^{-1}$.
Therefore the hadronic origin of the VHE emission is possible in the case of slow diffusion only.
Similar diffusion coefficients are also needed in other studies such as for the VHE emission in spatial coincidence with dense MCs around SNR W28 \citep{w28}.

\section{A compact nature}\label{compact}

\subsection{A faint PWN?}\label{pwn}

A likely scenario would be that both the X-ray and TeV sources stem from a PWN powered by a yet unknown pulsar.
Even if the non-thermal aspect of the X-ray emission is not well determined, its hard spectral index for the PL assumption is
indicative of an emission from the vicinity of a pulsar, e.g. magnetospheric or striped wind \citep[e.g.][]{Petri2007}. 
Therefore, despite the lack of observed pulsations in the object, a pulsar origin for \xmmu will be considered here.
The TeV emission would then be attributed to inverse Compton emission coming from the nebula powered by the putative pulsar.

It can be tested whether energetically a PWN scenario plausibly matches with the population of known TeV-emitting PWNe, under the hypothesis that the X-ray emission comes from the pulsar's magnetosphere.
The luminosity $L_{\rm X}(2-10\, \rm keV)$ of the X-ray point source can be translated to an estimate of the \.{E} of the hypothetical pulsar using the $L_{\rm X}/\dot{E}$ relation provided by \citet{li}.
The estimated spin-down luminosity is of the order of  $10^{37}\,\rm erg\, s^{-1}$ for a distance of 5 kpc, 
pointing towards a rather young pulsar age ($\lesssim \, 10^{5}\rm~years$).
Note that the compact size of the TeV source is also an indication for a fairly young object.
The \.{E}/d$^2$ for the same distance is $6 \times 10^{35} \rm erg \, s^{-1} kpc^{-2}$, 
corresponding to the band for which more than 40$\%$ of the PWNe are detected with \h \citep{Klepser}. 
Therefore, if the putative pulsar powers a TeV PWN, the latter should be detectable with the \h array.
A very similar conclusion is derived if the detected X-ray emission is assumed to stem from the hard core of the X-ray PWN.
The lack of an extended X-ray PWN around \xmmu could be attributed to yet insufficient statistics and \textit{XMM-Newton} angular resolution, 
or to the high absorption which would prevent any detection of the extension below ~3 keV (due to synchrotron cooling of the electrons, the outer region of the PWN would have a softer index than its core).
Another possibility would be that the X-ray PWN is underluminous \citep{Kargaltsev2008}.
Future high-resolution X-ray observations are thus needed to clarify this issue.
Together with the absence of detected X-ray pulsations, the PWN scenario is possible but remains for the moment still unconfirmed.

\subsection{A new binary system?}\label{binary}

The 2MASS catalog\footnote{http://www.ipac.caltech.edu/2mass/releases/allsky/} lists three infrared sources around the position of \xmmu within
the \textit{XMM-Newton} PSF of about 6$''$ (FWHM).
However only one faint source is located within the statistic positional error (2.3$''$) of \xmmu.
This source, \irmass, lies 1.9$''$ away from the position of the X-ray source. No optical counterpart is found, likely due to strong extinction in the Galactic plane.
The apparent magnitudes observed in the J, H, K bands are \mj, \mh, and \mk, respectively \citep{Skrutskie2006}.
The probability of a chance association between \irmass and \xmmu is around 2$\%$, following the approach by \citet{Akujor1987}.
To derive this value, all sources of the 2MASS catalog with $m_K \le 13$ in a 2$^\circ$ side box around \xmmu were selected, and the chance probability was computed to
detect one of them in a surface of 16.6 arcseconds$^2$ corresponding to the \textit{XMM-Newton} localisation error area.

The infrared source \irmass discovered in spatial coincidence with \xmmu
could suggest that the X-ray source resides in a binary system around a massive star.
Variable TeV emission from a number of gamma-ray binaries has already been detected \citep{Aharonian_1259,Aharonian_LS5039,Albert2009,Aliu2014}.
The optical brightness derived for \irmass is compatible with an association in a binary system with \xmmu if the measured X-ray absorption is mainly stemming from local gas around the X-ray source.
In the absence of orbitally modulated X-ray or TeV emission, the binary possibility remains unconfirmed, although the low chance probability association between the IR and X-ray sources 
seems to support this scenario.
The non-detection of variability could be either due to insufficient statistics or due to a specific geometrical shape of the binary system that would not produce modulated emission in gamma-rays.
Although one could expect strong GeV emission from gamma-ray binary systems, one of these objects has currently no GeV counterpart :  HESS J0632$+$057. Therefore the non-detection in GeV of \src does not rule out the binary scenario.
Moreover, HESS J0632$+$057 was unidentified at the time of its discovery \citep{HESS0632_2007} and its variability was only confirmed later on \citep{Hinton2009,Bongiorno2011}.
The similarities with HESS J0632$+$057 make \src a very good binary system candidate despite the absence of modulated emission.

\section{Conclusion}\label{conclusions}

Observations in the field of view of SNR \snr have led to the discovery of the point-like source \src lying on the edge of the 
SNR radio rim. Hadronic particles accelerated in the SNR \snr interacting with dense gas material could result in TeV emission through neutral pion production and decay in the case of slow CR diffusion.

On the other hand, a compelling X-ray counterpart, \xmmu, has been discovered, whose nature is yet to be established.
Together with the TeV emission and the infrared point source  \irmass, plausible object classifications are a young pulsar wind nebula or a gamma-ray binary.
Following the case of HESS J1943$+$213, an extragalactic origin such as an AGN could also be possible. However this scenario was disfavored due to the lack of GeV emission 
and point-like counterparts in radio data.

The TeV source properties strongly resemble the situation of HESS J0632$+$057 at the time of its discovery \citep{HESS0632_2007}, 
which only after extensive continued monitoring in X-rays and gamma-rays could be identified as a gamma-ray binary \citep{Hinton2009,Acciari2009,Bongiorno2011}. 
Point-like sources remain rare amongst all newly discovered VHE sources and \src is an excellent candidate for belonging to the rare and special class of gamma-ray binaries. 

Nevertheless, given the lack of a clear confirmation of the binary scenario through variability, other scenarios are also possible. 
The isolated PWN scenario, however, lacks an X-ray PWN detection despite XMM-Newton coverage, and the cosmic-ray-molecular cloud interaction scenario is hard to reconcile with the possible
association of \xmmu with \src. Further multiwavelength studies are therefore encouraged to establish (or ultimately reject) \src's classification as illuminated molecular cloud, gamma-ray binary or pulsar wind nebula.

\section*{Acknowledgements}
\small{
The support of the Namibian authorities and of the University of Namibia
in facilitating the construction and operation of H.E.S.S. is gratefully
acknowledged, as is the support by the German Ministry for Education and
Research (BMBF), the Max Planck Society, the French Ministry for Research,
the CNRS-IN2P3 and the Astroparticle Interdisciplinary Programme of the
CNRS, the U.K. Science and Technology Facilities Council (STFC),
the IPNP of the Charles University, the Czech Science Foundation, the Polish 
Ministry of Science and  Higher Education, the South African Department of
Science and Technology and National Research Foundation, and by the
University of Namibia. We appreciate the excellent work of the technical
support staff in Berlin, Durham, Hamburg, Heidelberg, Palaiseau, Paris,
Saclay, and in Namibia in the construction and operation of the
equipment.
}

This publication makes use of molecular line data from the Boston University-FCRAO Galactic Ring Survey (GRS). 
The GRS is a joint project of Boston University and Five College Radio Astronomy Observatory, funded by the National 
Science Foundation under grants AST-9800334, AST-0098562, $\&$ AST-0100793 

This publication makes use of data products from the Two Micron All Sky Survey, which is a joint project of the University 
of Massachusetts and the Infrared Processing and Analysis Center/California Institute of Technology, funded by the National Aeronautics and Space Administration and the National Science Foundation.

\bibliographystyle{mn2e_williams}

\bibliography{HESSJ1832-093_v16.5}
\label{lastpage}
\end{document}